\documentclass{elsarticle}

\AtBeginDocument{%
  \providecommand\BibTeX{{%
    \normalfont B\kern-0.5em{\scshape i\kern-0.25em b}\kern-0.8em\TeX}}}

\usepackage{tcolorbox}
\usepackage{booktabs}
\usepackage{ulem}
\usepackage{caption}
\usepackage{subcaption}
\usepackage{lineno,hyperref}
\modulolinenumbers[5]
\usepackage{rotating}
\usepackage{enumitem}
\setlist{nolistsep,leftmargin=*}
\definecolor{mygray}{gray}{0.6}

\usepackage{xpatch,environ,ragged2e}
\usepackage[colorinlistoftodos]{todonotes}
\RenewEnviron{abstract}{%
  \xdef\theabstracttext{%
    \unexpanded{%
      \def\baselinestretch{1}\noindent\unskip\textbf{Abstract}\par\medskip
      \noindent\unskip\ignorespaces}%
    \unexpanded\expandafter{\BODY}%
  }%
}
\def\theabstracttext{}
\xpatchcmd{\pprintMaketitle}
  {\ifvoid\absbox}
  {\ifx\theabstracttext\empty\else\printtheabstracttext\fi\ifvoid\absbox}
  {}{}
\newcommand{\printtheabstracttext}{{%
  \begin{trivlist}
  \normalfont\normalsize
  \item\relax
  \theabstracttext
  \end{trivlist}
}}

\usepackage{pgfplotstable}

\def\BibTeX{{\rm B\kern-.05em{\sc i\kern-.025em b}\kern-.08em
    T\kern-.1667em\lower.7ex\hbox{E}\kern-.125emX}}

\graphicspath{{pics/}}

\newcommand{\RqOne}{\textbf{\uline{RQ1. How many GitHub repositories link to academic papers?}}}

\newcommand{\RqOneOne}{\textit{RQ1.1 How many GitHub README.mds reference an academic paper?}}

\newcommand{\RqOneTwo}{\textit{RQ1.2 How publicly available are the academic papers referenced in GitHub README.mds?}}

\newcommand{\RqTwo}{\textbf{\uline{RQ2 What is the relationship between a GitHub repository and its referenced academic paper?}}}

\newcommand{\RqTwoOne}{\textit{RQ2.1 What kind of GitHub repository references an academic paper?}}

\newcommand{\RqTwoTwo}{\textit{RQ2.2 What is the affiliation of the owners of GitHub repositories that reference academic papers?}}

\newcommand{\RqTwoThree}{\textit{RQ2.3 Is there a direct relationship between the GitHub repository's contributors and its referenced academic paper?}}

\newcommand{\RqTwoFour}{\textit{RQ2.4 How do academic papers referenced in GitHub README.mds link back to the GitHub repository?}}

\newcommand{\RqTwoFive}{\textit{RQ2.5 How influential are academic papers that are referenced in GitHub repositories?}}

\newcommand{\RqTwoSix}{\textit{RQ2.6 How diverse is programming languages of the repositories that reference the academic paper?}}

\newcommand{\RqThree}{\textbf{\uline{RQ3 How does evolution affect the relationship between GitHub repository and academic paper?}}}

\newcommand{\RqThreeOne}{\textit{RQ3.1 Do academic papers get updated after being referenced in GitHub README.mds?}}

\newcommand{\RqThreeTwo}{\textit{RQ3.2 How do references to academic papers evolve in GitHub README.mds?}}

\begin{document}
\begin{sloppy}

\newcommand\rev[3]{\textcolor{red}{\sout{#1}} {\textcolor{blue}{#2}} {\todo[color=green!40]{\thesubsection{}. #3}}}



\title{GitHub Repositories with Links to Academic Papers: Public Access, Traceability, and Evolution}

\author[firstaddress]{Supatsara Wattanakriengkrai}
\ead{wattanakri.supatsara.ws3@is.naist.jp}
\author[firstaddress]{Bodin Chinthanet}
\ead{bodin.chinthanet.ay1@is.naist.jp}
\author[fourthaddress]{Hideaki Hata}
\ead{hata@shinshu-u.ac.jp}
\author[firstaddress]{Raula Gaikovina Kula}
\ead{raula-k@is.naist.jp}
\author[secondaddress]{Christoph Treude}
\ead{christoph.treude@adelaide.edu.au}
\author[thirdaddress]{Jin Guo}
\ead{jguo@cs.mcgill.ca}
\author[firstaddress]{Kenichi Matsumoto}
\ead{matumoto@is.naist.jp}

\address[firstaddress]{Nara Institute of Science and Technology, Japan}
\address[fourthaddress]{Shinshu University, Japan}
\address[secondaddress]{University of Adelaide, Australia}
\address[thirdaddress]{McGill University, Canada}

\begin{abstract}
Traceability between published scientific breakthroughs and their implementation is essential, especially in the case of open-source scientific software which implements bleeding-edge science in its code. However, aligning the link between GitHub repositories and academic papers can prove difficult, and the current practice of establishing and maintaining such links remains unknown.
This paper investigates the role of academic paper references contained in these repositories. We conduct a large-scale study of 20 thousand GitHub repositories that make references to academic papers. We use a mixed-methods approach to identify public access, traceability and evolutionary aspects of the links. Although referencing a paper is not typical, we find that a vast majority of referenced academic papers are public access. These repositories tend to be affiliated with academic communities. More than half of the papers do not link back to any repository. We find that academic papers from top-tier SE venues are not likely to reference a repository, but when they do, they usually link to a GitHub software repository. In a network of arXiv papers and referenced repositories, we find that the most referenced papers are (i) highly-cited in academia and (ii) are referenced by repositories written in different programming languages.
\end{abstract}

\maketitle
\section{Introduction and Motivation}
\label{sec:introduction}

Public access is one of the key concepts in open science, aiming to distribute research outputs (e.g., scientific academic papers) to the public to accelerate research~\cite{Woelfle2011OpenSI}, free of cost or other access barriers \cite{PeterSub3:online}. It has been adopted by increasingly more researchers and agencies across different domains. Meanwhile, software that is developed during the scientific research process is often released as open source software (OSS) to promote reproducible research and increase the chances of having a stronger impact. OSS can be copied and distributed at essentially no cost, potentially opening the door to unprecedented levels of sharing and collaborative innovation~\cite{howison2011}.

Reproducibility challenges and artifact tracks that have been hosted at international conferences have strengthened the connections between software and scientific papers. For example, consider the papers submitted to the Conference and Workshop on Neural Information Processing Systems (NeurIPS), one of the leading conferences for machine learning (ML) research. The number of papers containing the link to their source code have increased from 50\% to 75\% within one year (since 2018) \cite{ThisAIre54:online}.
It can be argued that a paper has a better chance for real impact if its code is accessible; the readers are able to examine the code, run the code, and assess the contributions considering full details of the work (which are often omitted in the paper).

Keeping up with the state of the art (SOTA) between software repositories and papers is always a challenge, especially due to the differences in cultures between scientists and software engineers~\cite{Segal_scientistsand}.
Initiatives such as \texttt{Papers With Code} \cite{PapersWi17:online} and \texttt{RedditSOTA} \cite{RedditSo59:online} are examples of maintaining traceability between two of the most prominent artifacts. More than advancing SOTA, traceability also increases the accountability of the scientific results. For example, it was found that over 100 published studies may have incorrect results due to a glitchy piece of Python code that was later discovered by researchers \cite{PythonCo27:online}.
In this example, the original authors were grateful that this error was successfully traced.

While the role of traceability between scientific artifacts can be important for the above reasons, there is no existing work investigating the current practice of creating and maintaining such trace links. 
In a related study of seven scientific repositories, \citet{MilewiczMSR19} found that the academic community (i.e., professors and students) was the main contributor to scientific repositories.
\citet{BraiekMSR18} studied ML research in open source development.
They found that academics and large corporations such as Google and Microsoft release ML frameworks under an open source license. 
Their study also revealed that companies are the main drivers, with hybrid teams comprising both engineers and professional scientists.
Although these papers are complementary, to the best of our knowledge, the link between repositories and papers has not been studied comprehensively.
We would like to understand how managing the public access, traceability and evolution of these links leads to potential opportunities and problems.
We follow the definition of traceability defined  by  CoEST  as  “the  ability  to  interrelate  any  uniquely  identifiable  software engineering artifact to any other maintain required links over time and use the resulting network to answer questions of both the software product and its development process” \cite{wwwcoest49:online}.  In our context, it also extends to using the resulting network between academic papers and software to advance scientific progress and to support knowledge transfer.
Evolution is defined as ``a paper update if a new version of the paper was published on arxiv.org or the paper was extended to a journal at a later point in time compared to when the paper is being referenced'' in our context.

In this paper, we lay the foundation for understanding the role of links to academic papers in GitHub repositories by collecting 20,278 GitHub repositories created between 2014 and 2018.
Our preliminary analysis shows that links come from the README.md, a significant documentation resource for scientific software.
It is important to understand developers' typical knowledge sharing activities by referencing external sources to improve software documentation in practice.
Results show that referencing academic papers in README.mds is not frequent in GitHub repositories
and that the vast majority of academic papers that are referenced from GitHub repositories are public access.
In terms of relationship and traceability between GitHub repository and academic paper, machine learning is the most frequent topic while individuals that are affiliated with academic communities (i.e., universities) tend to own these repositories.
More than half of the referenced papers do not link back to any repository.
In a deeper look at repositories that reference arXiv papers, we find that most referenced papers are high-impact, influential, and do align with academia, being referenced by repositories written in different programming languages.
Finally, the evolution is slow, with little change to academic papers that are referenced in GitHub README.mds.
These changes are likely not reflected in the GitHub repositories which reference them since updates to paper links are rare.

This paper's contributions are three-fold: (i) a large-scale and comprehensive study of 20 thousand links to establish the frequency of links from GitHub repositories to academic papers, (ii) a mixed-methods study to identify public access, traceability, and evolutionary aspects of such links and (iii) availability of an online appendix, which contains our qualitative coding results in this study.

\section{Research Method}
\label{sec:research_method}
In this section, we present our research questions, data collection methodology, and we introduce the data contained in our online appendix.

\subsection{Research Questions}
\label{subsec:research_questions}
The main goal of our study is to gain insights into the links between software repositories on GitHub and academic papers, and to uncover potential issues related to public access, traceability, and evolution.
Based on this goal, we formulated three main research questions to guide our study.
We now present each of these questions, along with their motivation.
\\
\begin{description}[style=unboxed,leftmargin=0cm]
\item \noindent {\RqOne~}
\end{description}
The motivation of RQ1 is to understand how often GitHub repositories contain salient references to academic papers (i.e., in their documentation files).
We then breakdown this question into two sub-questions, based on our preliminary analysis of the file types in GitHub repositories that contain such references.
Complementary to \cite{10.1109/ICSE.2019.00123} and initiatives such as \texttt{Papers With Code} \cite{PapersWi17:online} and \texttt{RedditSOTA} \cite{RedditSo59:online}, we would like to perform a deeper analysis to evaluate whether or not links from GitHub README.md to academic papers are common. This leads to the first sub-question, \RqOneOne

Our second sub-question (\textit{RQ1.2}) is related to how often these links are public access. 
Answering this research question will provide us with insights on the current state of public access with regard to academic papers referenced from GitHub repositories -- if they are hidden behind paywalls, researchers and practitioners will find it difficult to access them.
Thus, our second sub-question is as follows: 
\RqOneTwo
\\
\begin{description}[style=unboxed,leftmargin=0cm]
\item \noindent \RqTwo~
\end{description}
The motivation of RQ2 is to take a closer look at the characteristics of GitHub repositories and the papers they reference. 
We analyze this relationship from six different aspects.
First, we would like to understand the domain of those GitHub repositories that reference academic papers.
Answering this research question will provide us with the characteristic of GitHub repositories, especially the type of scientific artifact that references these academic papers.
This leads us to the first sub-question, \RqTwoOne

The motivation of our second sub-question (\textit{RQ2.2}) is to examine the affiliation of the main contributors, to see whether they are from industry or academia.
This is an expansion from \citet{MilewiczMSR19}, which only studied seven GitHub repositories.
This leads us to the second sub-question, \RqTwoTwo

Extending from the second sub-question, the third sub-question (\textit{RQ2.3}) explores whether or not the repository contributors and the paper authors are the same. 
Answering this question should allow us to identify the extent to which the paper is being implemented by developers that are not co-authors of the paper.
This leads us to the third sub-question, \RqTwoThree

In regards to the forth sub-question, (\textit{RQ2.4}) we explore whether the academic papers link back to the repository in which we originally found them.
We assume that papers that link back to the repository acknowledge the existence of the repository.
This leads us to the fourth sub-question, \RqTwoFour

Finally, our motivation for the fifth and sixth sub-questions, (\textit{RQ2.5} and \textit{RQ2.6}) is to look into the impact of the papers that are being referenced. 
As highlighted by \citet{ThisAIre54:online},
answering this research question can shed light on whether referenced papers have high impact in terms of citations and also the programming languages of the implementations of the paper.
This leads us to the fifth and the sixth sub-question, 
\RqTwoFive~and \RqTwoSix
\\
\begin{description}[style=unboxed,leftmargin=0cm]
\item \noindent \RqThree~
\end{description}
Our aim for RQ3 is to investigate the relationship between GitHub software repositories and the referenced academic papers from an evolutionary standpoint.
Following the report regarding the error discovered by researchers \cite{PythonCo27:online} that affected over 100 published studies, it is important that both the owners of the repositories and the paper co-authors are aware of changes in the links.
We breakdown this question into two sub-questions.
In response to our first sub-question, (\textit{RQ3.1}) we would like to understand whether the academic papers referenced from software repositories get updated.
This leads us to the first sub-question, \RqThreeOne

In response to our second sub-question, (\textit{RQ3.2}) we would like to understand whether the repository owner is aware of when there is a new version of the referenced academic paper. 
This could be the case when links to academic papers or their metadata are evolved on GitHub.
This leads us to the second sub-question, \RqThreeTwo

\subsection{Data Preparation}
\label{subsec:data_collection}
We now describe our method for the preparation of our target GitHub repositories and identification of referencing academic papers in software documentation files. 

\paragraph{Candidate Repositories}
To cover a wider range of software repositories, similar to previous work~\cite{10.1109/ICSE.2019.00123}, we chose target repositories that are being hosted on GitHub written in popular programming languages (i.e., C, C++, Java, JavaScript, Python, PHP, and Ruby).
To ensure that we covered mature projects that were active for at least 2 years, i.e., having at least 100 commits in the most active two years, we mined repositories created from 2014 to 2018.
Using GHTorrent \cite{Download10:online, 10.5555/2487085.2487132}, we extracted more than ten million candidate repositories.
\begin{table}
\caption{Targeted GitHub repositories}
\label{tab:collected_repo}
\centering
\scalebox{0.70}{
\begin{tabular}{@{}lrrrr@{}}
\toprule
\multicolumn{1}{c}{} & \multicolumn{1}{c}{\textbf{\# candidate}} & \multicolumn{1}{c}{\textbf{\# obtained}} & \multicolumn{2}{c}{\textbf{\# pattern matched}}
\\ \midrule
C & 501,226 & 222,614 & 876 & (0.4\%) \\
C++ & 700,786 & 329,201 & 3,263 & (1.0\%) \\ 
Java & 2,627,220 & 935,738 & 1,213 & (0.1\%) \\ 
JavaScript & 3,182,589 & 1,689,126 & 687 & (0.04\%) \\ 
Python & 1,587,765 & 818,079 & 14,073 & (1.7\%) \\ 
PHP & 951,202 & 415,039 & 57 & (0.01\%) \\ 
Ruby & 792,523 & 395,669 & 109 & (0.03\%) \\ 
\midrule
\textbf{sum} & \textbf{10,343,311} & \textbf{4,805,466} & \textbf{20,278}  & \textbf{(0.4\%)} \\ 
\bottomrule
\end{tabular}
}
\end{table}

\paragraph{Preliminary Analysis: Which file in a GitHub repository is the place where a developer is most likely to reference an academic paper?}
Our initial explorations showed that reference extraction is not trivial and can be prone to many false positives. As a preliminary study, we conducted an exploration of (i) different search string matching patterns and (ii) the most feasible location for the reference of an academic paper.

\begin{table}[]
\caption{The motivation for choosing search string matching patterns is that they are more likely to be links to academic papers, with the a low false positive rate. There are 20,278 README.mds that match the search strings.}
\centering
\label{tab:our-patterns}
\scalebox{0.70}
{
\begin{tabular}{@{}lr@{}}
\toprule
Match Pattern            & \multicolumn{1}{c}{\textbf{search string}} \\ \midrule
explicit DOI                                                    &    \texttt{`doi$\backslash$s*=$\backslash$s*[\{"]'}, \texttt{`//doi$\backslash$.'}, \texttt{`dx$\backslash$.doi$\backslash$.'}                                                  \\
explicit \BibTeX{}                                             &    \texttt{`@article\{'}, \texttt{`@inproceedings\{'}, \texttt{`@misc\{'}                                                 \\ 
explicit ArXiv                                                       &        \texttt{`arxiv.org'}                                             \\ \bottomrule
\end{tabular}
}
\end{table}

\begin{table}[]
\caption{Analysis of README.mds with expanded search string patterns.}
\label{tab:other-patterns}
\centering
\scalebox{0.70}
{
\begin{tabular}{@{}lrrr@{}}
\toprule
Expand Pattern  & \multicolumn{1}{c}{\textbf{\# Expand Only}} & \multicolumn{1}{c}{\textbf{\# Match $\cap$ Expand}} & \multicolumn{1}{l}{\textbf{\% not papers (FP)}} \\ \midrule
PDF extension                 & 9,880                                                         & 866                                                          & 76\% (38/50)                                                             \\
All \BibTeX{}             & 154                                                         & 161                                                           & 16\%  (8/50)                                                            \\
other patterns & 3,734                                                         & 1,041                                                           & 8\% (4/50)                                                              \\ \bottomrule
\end{tabular}
}
\end{table}

For the first goal,
we sampled 500 repositories from each languages and then experimented with different search string patterns. After multiple iterations, we ended up with seven search strings using \texttt{grep -riIG}, covering the explicit arXiv, DOI, and \BibTeX{} formats.
Since our search strings exactly match references in explicit \BibTeX{} formats, and explicit links pointing to DOI for academic papers and arXiv.org as shown in Table \ref{tab:our-patterns}, we are confident that the search strings are able to detect links to academic papers, with the lowest change for a false positive.

We characterize the bias of our matching patterns to compare the coverage of our patterns against a more expanded pattern and report false positives.
To do so, we relaxed our matching pattern rules to expand the search space (i.e., (PDF extension) included all links that had a PDF extension, (All \BibTeX{}) all \BibTeX{} formatting, (other patterns) other publishing sources such as IEEE, ACM, and PeerJ). To validate whether the result is a link to an academic paper, the first author manually checked 50 random samples of each expanded pattern.
Table \ref{tab:other-patterns} shows our results. First, we see that although the PDF extension returns a high number of results (9,880), it also has a high false-positive rate (76\%). Second, using all \BibTeX{} does not return as many results (154). The preliminary results also suggest that future work should investigate the use of the other patterns which achieved better performance but which also contain a number of hard-coded URLs and might therefore not generalize well.

For the second goal, we used the same 500 repositories and applied our search string over all files in the repositories. We were able to extract 47 repositories and 196 different files that returned at least one result. From these 196 files, a co-author manually excluded links that matched the following criteria: (a) is a link from a cloned/duplicated file, (b) is a link to a dataset, (c) is a link which is part of a dataset.

After applying this criteria, we end up with 72 valid references to an academic paper.
The 72 files comprised of the following file types:  (1) .py: 14, (2) .md: 11, (3) .h: 10, (4) .c: 7, and (5) .bib: 6. Although references in Python (.py: 14) source code are the most frequent, Python repositories would only cover 17\% (i.e., only 800K out of 4.8M repositories) of our collected repositories. Furthermore, related work \cite{10.1109/ICSE.2019.00123,2019arXiv191006932I} shows that only up to 3\% of source code links are references to academic papers. As a result of these considerations, we only consider documentation files (i.e., .md files) as the next highest. A closer investigation of the .md files showed that README.md is the most common file, i.e., covering 9 out of 11 files. Another reason to choose README.mds for our study is the fact that these files build the facade for a repository, i.e., they are the first thing a user would see when looking at a repository. Information in README.mds also likely covers the entire repository, not just a specific aspect of the source code (as might be the case in .py files).

\begin{tcolorbox}
\textbf{Summary}: 
Preliminary analysis shows papers are being referenced in both source code and documentation files, with .py files being the most frequent. However, the README.md is the next most frequent location where developers make references to an academic paper across software repositories of difference languages.
\end{tcolorbox}

\subsection{Sample Data Creation}
\label{ssec:sampling}
As shown in Table \ref{tab:collected_repo}, we proceeded to collect 20,278 README.mds from 4.8 million README.mds from our candidate repositories.
Then by applying our search string from the preliminary study, we obtained more than 20 thousand files in this step as shown in Table~\ref{tab:collected_repo}.
Some of our research questions require a representative sample for a deeper analysis. 
From the 20,278 README.mds, we then drew a statistically representative random sample.
It is important to note that the qualitative results are not mutually exclusive.
The required sample size was 377, which was calculated so that our conclusions about the ratio of files with a specific characteristic would generalize to all files with a confidence level of 95\% and a confidence interval of 5\% \cite{SampleSi17:online}.
To answer all qualitative research questions, we use the following coding approach, which includes four authors to first independently code a subset of 30 files.
Then to check agreement for qualitative analyses, a kappa agreement score (or Cohen's Kappa) is calculated, comparing the coding between all four authors.
The kappa agreement varies from 0 to 1 where 0 is no agreement and 1 is perfect agreement.
This process is iterated until there is a kappa agreement of more than 0.8 or ``almost perfect''~\cite{viera2005understanding}.
Once this is achieved, the remaining data is coded by a single author.
The appendix of all our qualitative coding results, also with all disagreement cases from our qualitative investigation is available at \url{https://github.com/NAIST-SE/GH2Papers}.

\section{Findings}
\label{sec:findings}

\subsection{Frequency and Public Access (RQ1)} 
To understand the frequency and public access of academic papers that are referenced in repositories (RQ1), we examined the existence of references to academic papers in the sample of 377 README.mds (RQ1.1) and report the result of our qualitative analysis with 344 academic papers referenced in README.mds for public access (RQ1.2). 

\begin{figure}[t]
\centering
\begin{footnotesize}
\begin{tikzpicture}[scale=0.9]
\begin{axis}[
    align=center,
    legend pos=south east,
    y=0.5cm,
    x=0.02cm,
    enlarge y limits={abs=0.25cm},
    symbolic y coords={no reference, reference to data, reference to paper},
    axis line style={opacity=0},
    major tick style={draw=none},
    ytick=data,
    xmin = 0,
    xlabel = \# GitHub repositories,
    nodes near coords,
    nodes near coords align={horizontal},
    point meta=rawx
    ]
    \addplot[xbar,fill=gray,draw=none] coordinates {
        (344,reference to paper)
        (18,reference to data)
        (15,no reference)
    };
\end{axis}
\end{tikzpicture}
\end{footnotesize}
\caption{Answering RQ1.1, we show that up to 344 of the 377 READMEs reference academic papers.}
\label{fig:sample}
\end{figure}
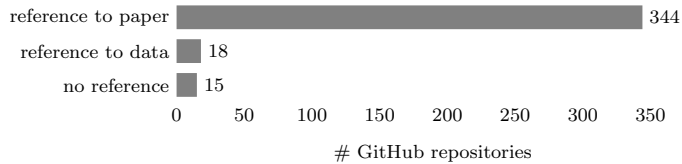

\paragraph{(RQ1.1) README.md referencing Academic Papers}
With more than 20,000 README.md references to academic papers, there is enough evidence to suggest that references in README.mds is a phenomenon worthy of further exploration. As mentioned in our approach for the qualitative study, we then analyze a representative sample of the dataset to understand the kinds of references to academic papers. In cases where multiple academic papers are referenced, we only select one representative academic paper per README.md. This is based on whether the repository and paper topics align with each other. In rare cases, if there are multiple related references that all appear to be close, the first paper was selected.
The kappa agreement among the four authors was 0.96 (interpreted as ``almost perfect''~\cite{viera2005understanding}). We used the following three codes to annotate referencing in the README.mds.

\begin{itemize}
\item \textit{reference to paper}: the README.md is referencing an academic paper.
\item \textit{reference to data}: the README.md is referencing a data repository.
\item \textit{no reference}: the README.md has no reference to an academic paper or a data repository. 
\end{itemize}

As shown in Figure~\ref{fig:sample}, the majority of the README.mds (91\%, 344/377) contain a link to an academic paper.
We observed that 5\% (18/377) of the files have links to data repositories instead of academic papers, mainly with the patterns related to DOI (e.g., because of a redirect to Figshare).
Our search string detected the remaining 15 repositories as having no reference, mainly because they listed a \BibTeX{} format for anyone to cite that repository.
In the end, from all 344 README.mds actually referencing academic papers, we observed that the majority of representative academic papers are available on arXiv.org \cite{arXivorg85:online} (66\%, 226/344), followed by ieeexplore.ieee.org (3\%, 11/344), researchgate.net (3\%, 11/344), sciencedirect.com (2.9\%, 10/344), academic.oup.com (2.3\%, 8/344), link.springer.com (1.7\%, 6/344), biorxiv.org (1.5\%, 5/344), and nature.com (1.5\%, 5/344). The appendix of the results is available online (see Section 2.3).  \textit{To answer RQ1.1, our findings confirm the types of academic papers get referenced in README.mds.}

\paragraph{(RQ1.2) Public Access}

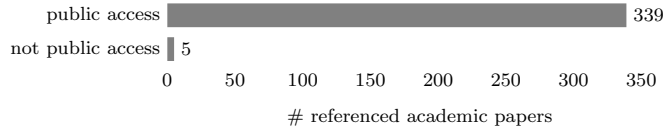
\begin{figure}
\centering
\begin{footnotesize}
\begin{tikzpicture}[scale=0.9]
\begin{axis}[
    align=center,
    y=0.5cm,
    x=0.02cm,
    enlarge y limits={abs=0.25cm},
    symbolic y coords={not public access, public access},
    axis line style={opacity=0},
    major tick style={draw=none},
    xlabel = \# referenced academic papers,
    ytick=data,
    xmin = 0,
    nodes near coords,
    nodes near coords align={horizontal},
    point meta=rawx
]
\addplot[xbar,fill=gray,draw=none] coordinates {
    (339,public access)
    (5,not public access)
};
\end{axis}
\end{tikzpicture}
\end{footnotesize}
\caption{Answering RQ1.2, we show that the majority of academic papers referenced in GitHub repositories are public access.}
\label{fig:rq1.2}
\end{figure}

Since our main goal is to better understand the links between GitHub repositories and academic papers, we particularly investigate 344 files with actual academic paper references for the subsequent qualitative analyses.
For our coding method, four authors individually investigated the aforementioned 30 README.mds, and listed the URL links to representative papers.
If there is no URL reference (e.g., \BibTeX{} entry without URLs) in the README.mds, we manually searched for an accessible version of the referenced papers.
We obtained a kappa agreement of 0.92 or ``almost perfect''~\cite{viera2005understanding}, and the remaining cases were investigated by a single author.

To investigate whether the referenced academic papers are accessible without any subscription or payment, we manually investigated each of the academic papers referenced in the README.mds in our sample.
The kappa agreement was 1.0 or ``perfect'' agreement~\cite{viera2005understanding}.
We use the following two codes to indicate whether the papers referenced in the README.md are publicly available:

\begin{itemize}
\item \textit{public access}: the referenced academic paper is available online including authors' preprint.
\item \textit{not public access}: the referenced academic paper is not public access, and authors do not provide their preprint.
\end{itemize}

Our result is very encouraging as seen in Figure~\ref{fig:rq1.2}. For the academic papers that are referenced from GitHub repositories, the vast majority are publicly available, and only five (1.5\%) of the academic papers are not public access yet (i.e., readers need to pay for the papers).
To answer RQ1.2, our findings indicate that \textit{more than 98\% of academic papers referenced in the GitHub repositories are available to the public.} 

\begin{tcolorbox}
\textbf{Summary}: 
With more than 20,000 GitHub repositories that reference an academic paper, the phenomenon is worthy of further exploration. We manually validate that the vast majority of academic papers referenced from GitHub repositories are public access.
\end{tcolorbox}

\subsection{Relationship and Traceability (RQ2)}
To understand the relationship between GitHub repositories and the referenced academic papers, and the traceability of the referenced papers, we conducted qualitative analyses on 339 repositories that reference public access academic papers (RQ2.1 - RQ2.4), a case study of 20,278 repositories whose README.mds matched our patterns (RQ2.1, RQ2.5, and RQ2.6), and a case study of a sample of 2,032 academic papers from the top 7 software engineering venues (RQ2.4). 

 \begin{figure}[t]
        \centering
        \begin{footnotesize}
        \begin{tikzpicture}[scale=0.9]
        \begin{axis}[
            align=center,
            y=0.5cm,
            x=0.05cm,
            enlarge y limits={abs=0.25cm},
            symbolic y coords={other, sensors, quantum, networks, robotics, natural language processing, other machine learning, computer vision, deep learning},
            axis line style={opacity=0},
            major tick style={draw=none},
            ytick=data,
            xmin = 0,
            xlabel = \# GitHub repositories,
            nodes near coords,
            nodes near coords align={horizontal},
            point meta=rawx
        ]
        \addplot[xbar,fill=gray,draw=none] coordinates {
            (92,deep learning)
            (83,other)
            (72,computer vision)
            (40,other machine learning)
            (29,natural language processing)
            (8,robotics)
            (7,networks)
            (6,quantum)
            (2,sensors)
        };
        \end{axis}
        \end{tikzpicture}
        \end{footnotesize}
        \caption{Answering RQ2.1, we show that the references to papers are frequently made in machine learning-related GitHub repositories.}
        \label{fig:rq2.1}
    \end{figure}
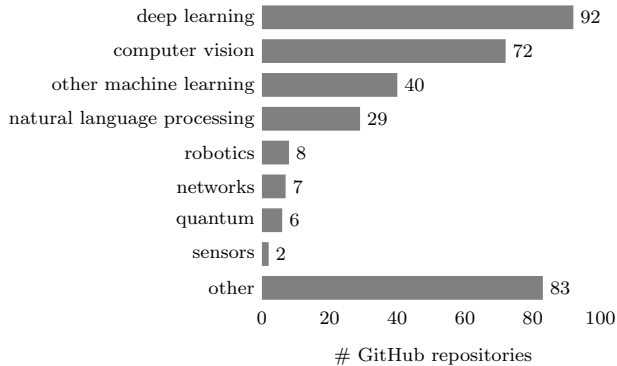

\paragraph{(RQ2.1) Characteristics of GitHub Repositories}
First, we investigated software domains of GitHub repositories with a qualitative analysis.
The same 30 cases were independently coded by the four authors, similar to Section~\ref{ssec:sampling}. We discussed our free-form codes and derived the following nine codes. Using the new coding guide, all four authors then independently re-coded the 30 GitHub repositories. In case a repository can be tagged with multiple codes (e.g., a computer vision repository uses deep learning), we consider the main purpose of the repository described in the README.md to determine the code. 
The kappa agreement was 0.83 or ``almost perfect''~\cite{viera2005understanding}. Based on this encouraging agreement value, the remaining cases were then coded by a single author.
\begin{itemize}
\item \textit{deep learning}: the repository is specifically focused on the deep neural network architecture.
\item \textit{computer vision}: the repository implements algorithms that solve computer vision tasks such as analyzing images or videos for object recognition and tracking.
\item \textit{natural language processing}: the repository is concerned with the processing and/or understanding of human natural languages (e.g., English).
\item \textit{other machine learning}: the repository is related to machine learning, but does not fit any of the previous three codes.
\item \textit{quantum}: the repository is related to quantum computing.
\item \textit{robotics}: the repository is related to robotics, but not from a machine learning or computer vision aspect.
\item \textit{networks}: the repository is related to analyzing the properties of network structures and their applications.
\item \textit{sensors}: the repository is related to sensor technologies.
\item \textit{other}: anything that does not fit any of the previous codes.
\end{itemize}

As shown in Figure~\ref{fig:rq2.1}, 
we found that the most common software domain of the GitHub repositories that reference academic papers is ``deep learning'', followed by ``computer vision'' and ``other machine learning''.
In other words, machine learning is the most common, covering three quarters of the repositories in our sample.
Also, more than 20\% of the GitHub repositories referencing academic papers belong to the ``other'' code, for instance, web API, biology, or chemistry. This indicates the diversity of these GitHub repositories.

\begin{figure}[t]
\centering
    \begin{subfigure}[]{0.45\textwidth}
    \centering
    \includegraphics[width=\textwidth]{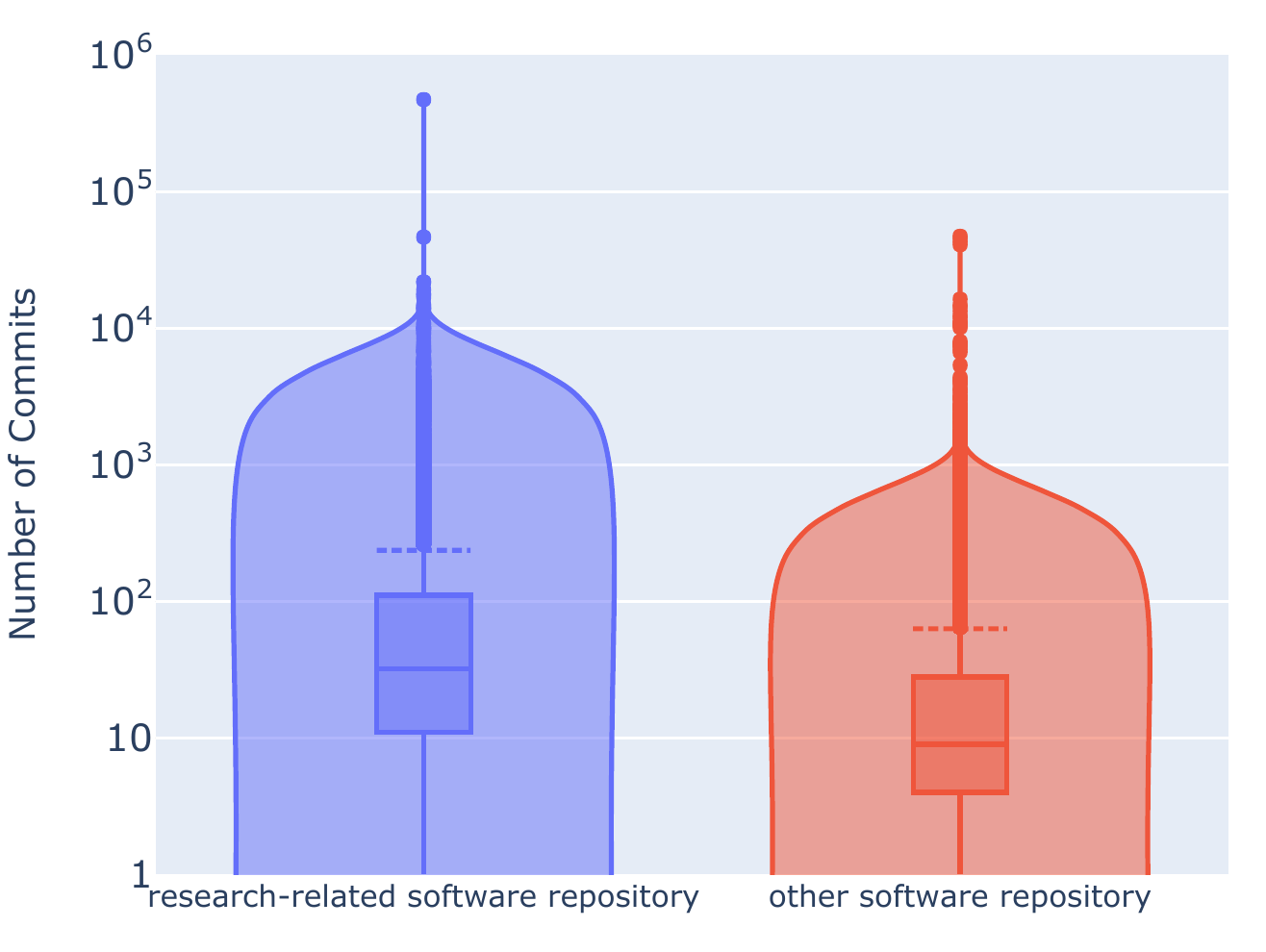}
    \caption{The number of commits}
    \label{fig:compare_commits_violin}
    \end{subfigure}
    \hfill
    \begin{subfigure}[]{0.45\textwidth}
    \centering
    \includegraphics[width=\textwidth]{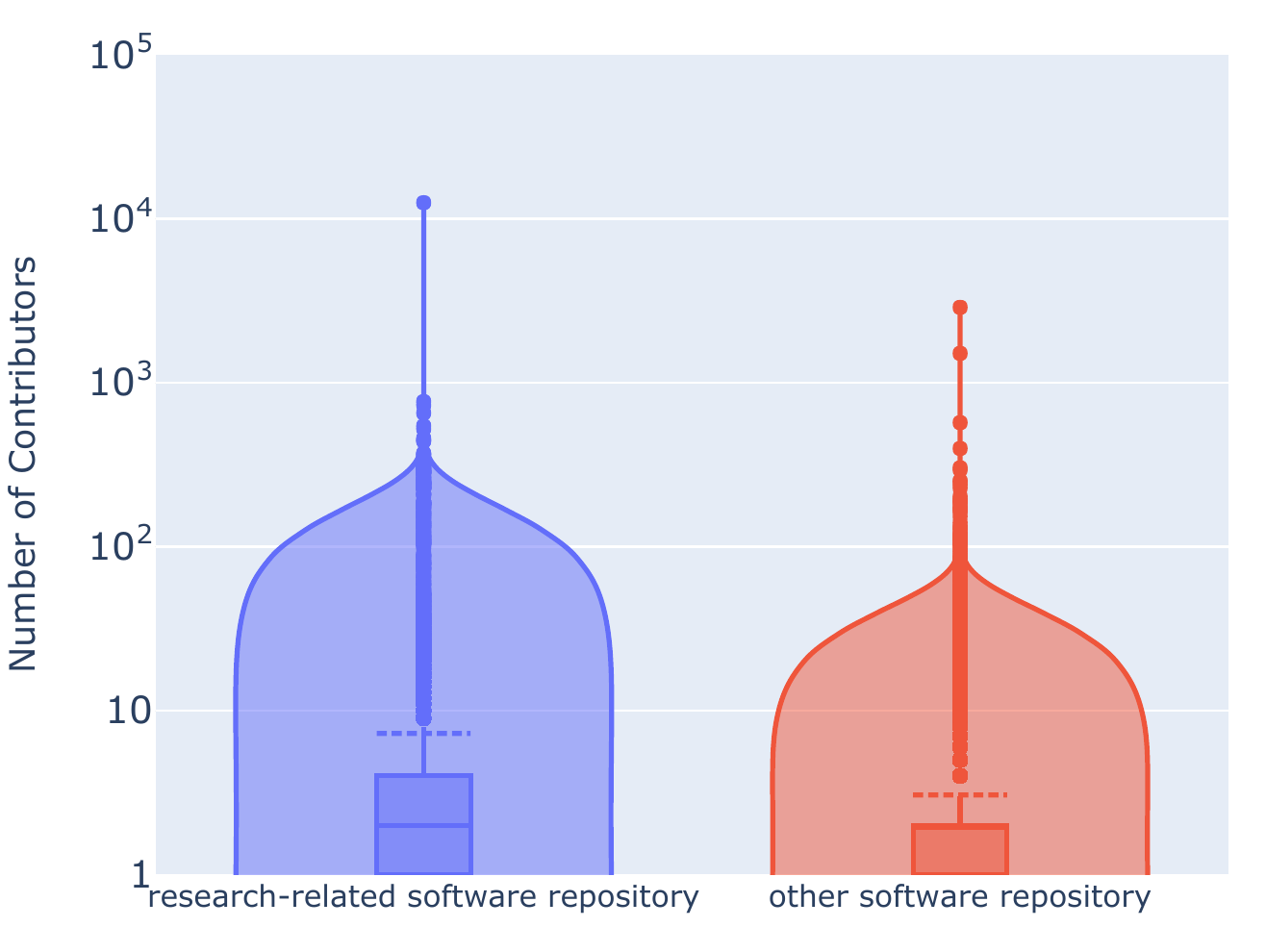}
    \caption{The number of contributors}
    \label{fig:compare_contributors_violin}
    \end{subfigure}
    \caption{Comparison between research-related software repositories linked from GitHub README.mds and other software repositories}
    \label{fig:rq2.1-compare}
\end{figure}

To complement the result of the above qualitative analysis, we also investigated the characteristics of GitHub repositories 
in terms of development activity (through number of commits) and personnel (through number of contributors).
To do so, we first obtained a list of GitHub repositories and their corresponding information, e.g., programming language, from GHTorrent.
Then, we downloaded README.mds of the repositories using a GitHub API \footnote{\url{https://docs.github.com/en/rest}}.
Finally, we attempted to clone 20,278 GitHub repositories whose README.mds match the patterns.
To compare with these repositories, we also attempted to clone 20,000 GitHub repositories which did not match the patterns (a random sample of the 4,805,466 repositories excluding the matching repositories).
Unfortunately, some repositories could not be cloned because of several technical errors.
As a result, we obtained 20,254 GitHub repositories matching the patterns and 19,932 GitHub repositories that did not match the patterns.
Figure \ref{fig:rq2.1-compare} shows the comparison between research-related software repositories and other software repositories, i.e., GitHub repositories that matched our patterns and other GitHub repositories having README.mds which did not match the patterns. 
As shown in Figure \ref{fig:compare_commits_violin}, we found that research-related software repositories have the number of commits higher than other software repositories ($\bar{x} = 32, \mu = 237.11$ for research-related repositories and $\bar{x} = 9, \mu = 63.05$ for other repositories). 
According to \citet{10.1145/2597073.2597074}, we suspect that some possible reasons for the lower number of commits of the non-research-related software repositories are that most GitHub repositories are inactive, personal, not a project, or not for software development, and many active projects do not conduct all their software development in GitHub.
Figure \ref{fig:compare_contributors_violin} also shows that research-related repositories have more contributors ($\bar{x} = 2, \mu = 7.26$ for research-related repositories and $\bar{x} = 2, \mu = 3.06$ for other software repositories).
To confirm that the differences between the two types of repositories are statistically significant, we use a Wilcoxon Rank-Sum test \cite{mann1947}. 
The result shows that the differences between both types of repositories are statistically significant in both commits and contributors aspects, i.e., \textit{p-value} $<$ 0.001.

To answer RQ2.1, our findings indicate that \textit{machine learning-related repositories, especially deep learning, are the most frequent among GitHub repositories with links to academic papers. These GitHub repositories have the number of commits and contributors significantly higher than general software repositories in GitHub.}

\paragraph{(RQ2.2) Affiliation}
The following list shows four codes that emerged from our analysis regarding repository affiliation, along with a short description. In all cases, we attempted to determine the affiliation of the repository owner through their GitHub profile page, either by reading information directly on the profile page or by following links.
The kappa agreement was 0.84 or ``almost perfect''~\cite{viera2005understanding}.

\begin{itemize}
\item \textit{university}: the owners are affiliated with universities.
\item \textit{industry}: the owners are affiliated with companies.
\item \textit{both}: the owners are affiliated with universities and companies at the time the academic paper is being referenced in the GitHub README.md.
\item \textit{unknown}: we cannot determine the affiliation from the profile of the owner of the GitHub repository.
\end{itemize}

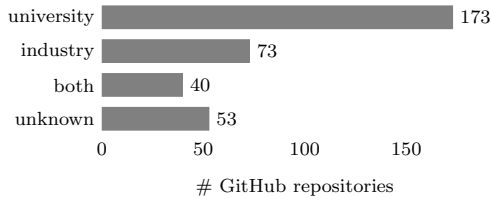
\begin{figure}[t]
        \centering
        \begin{footnotesize}
        \begin{tikzpicture}[scale=0.9]
        \begin{axis}[
            align=center,
            y=0.5cm,
            x=0.03cm,
            enlarge y limits={abs=0.25cm},
            symbolic y coords={unknown, both, industry, university},
            axis line style={opacity=0},
            major tick style={draw=none},
            ytick=data,
            xmin = 0,
            xlabel = \# GitHub repositories,
            nodes near coords,
            nodes near coords align={horizontal},
            point meta=rawx
        ]
        \addplot[xbar,fill=gray,draw=none] coordinates {
            (173,university)
            (73,industry)
            (53,unknown)
            (40,both)
        };
        \end{axis}
        \end{tikzpicture}
        \end{footnotesize}
        \caption{Answering RQ2.2, we show that the GitHub repositories which reference academic papers are commonly affiliated with universities.}
        \label{fig:rq2.2}
\end{figure}

Figure \ref{fig:rq2.2} shows the frequency of the different types of affiliations of GitHub repository owners from our statistically representative sample.
Unsurprisingly, the most common type of affiliation is ``university'', accounting for more than half of the representative repositories, followed by ``industry'' (e.g., companies and research centers), which is accounting for 21\% of the repositories from our sample.
To answer RQ2.2, our findings indicate that \textit{the majority of the GitHub repositories which reference academic papers are affiliated with universities.}  

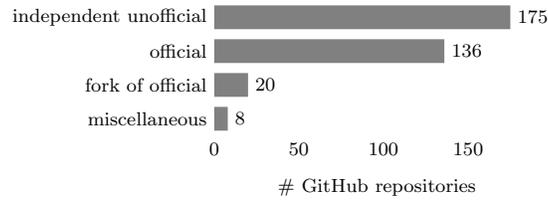
\begin{figure}[]
        \centering
        \begin{footnotesize}
        \begin{tikzpicture}[scale=0.9]
        \begin{axis}[
            align=center,
            y=0.5cm,
            x=0.025cm,
            enlarge y limits={abs=0.25cm},
            symbolic y coords={miscellaneous,fork of official,official,independent unofficial},
            axis line style={opacity=0},
            major tick style={draw=none},
            ytick=data,
            xmin = 0,
            xlabel = \# GitHub repositories,
            nodes near coords,
            nodes near coords align={horizontal},
            point meta=rawx
        ]
        \addplot[xbar,fill=gray,draw=none] coordinates {
            (175,independent unofficial)
            (136,official)
            (20,fork of official)
            (8,miscellaneous)
        };
\textbf{}        \end{axis}
        \end{tikzpicture}
        \end{footnotesize}
        \caption{Answering RQ2.3, we show that the large number of referenced academic papers (40\%) in GitHub repositories are referenced by the paper authors.}
        \label{fig:rq2.3}
    \end{figure}

\paragraph{(RQ2.3) Relationship between Repositories and Papers}
To answer this question, we explored the context of the references, and considered any information in the repositories and corresponding papers if it was necessary.
The following list shows four codes that emerged from our analysis of the relationship between the GitHub repositories and the academic papers that these repositories are referencing.
The kappa agreement was 0.89 or ``almost perfect''~\cite{viera2005understanding}.

\begin{itemize}
\item \textit{official}: the owner of the software repository is one of the authors of the referenced academic paper.
\item \textit{fork of official}: the owner of the software repository is not one of the authors of the referenced academic paper, and the target repository is forked from the official repository that belongs to one of the authors of the referenced paper.
\item \textit{independent unofficial}: the owner of the software repository is not one of the authors of the referenced academic paper, and there is no obvious relationship between the target repository and the official repository of the paper, if it 
exists.
\item \textit{miscellaneous}: the owner of the non-software repository is not one of the authors of the referenced academic paper.
\end{itemize}

Figure \ref{fig:rq2.3} shows the results of our qualitative analysis.
We found that in more than 50\% cases, the relationship is not an official one, i.e., the owners of the repositories had implemented their software based on academic papers that do not belong to the owners. In contrast, 40\% of the repository owners in our sample are referencing their own academic work. To answer RQ2.3, our findings indicate that \textit{most of the references to academic papers in GitHub repositories are referenced by repository owners other than paper authors.}

\begin{tcolorbox}
\textbf{Summary}: 
GitHub repositories tend to be closely related to the academic paper that they are referencing. Results indicate that GitHub repositories that discuss machine learning topics in their README.md tend to reference an academic paper, with individuals that are affiliated with academic communities (i.e., universities) owning these repositories.
\end{tcolorbox}

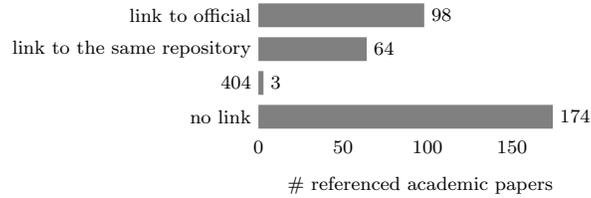
\begin{figure}[t]
        \centering
        \begin{footnotesize}
        \begin{tikzpicture}[scale=0.9]
        \begin{axis}[
            align=center,
            y=0.5cm,
            x=0.025cm,
            enlarge y limits={abs=0.25cm},
            symbolic y coords={no link,404,link to the same repository,link to official},
            axis line style={opacity=0},
            major tick style={draw=none},
            ytick=data,
            xlabel = \# referenced academic papers,
            xmin = 0,
            nodes near coords,
            nodes near coords align={horizontal},
            point meta=rawx
        ]
        \addplot[xbar,fill=gray,draw=none] coordinates {
            (98,link to official)
            (64,link to the same repository)
            (3,404)
            (174,no link)
        };
        \end{axis}
        \end{tikzpicture}
        \end{footnotesize}
        \caption{Answering RQ2.4, we show that more than half of academic papers do not reference back to the GitHub repositories.}
        \label{fig:rq2.4}
    \end{figure}

\paragraph{(RQ2.4) Paper References back to the Repository}
The following list shows four codes that emerged from our analysis of potential bi-directional links, i.e., links back from the referenced paper to the GitHub repository.
The kappa agreement was 0.91 or ``almost perfect''~\cite{viera2005understanding}.

\begin{itemize}
\item \textit{link to official}: the referenced paper has a link to the official repository, which is different from the GitHub repository that contains a reference to that paper.
\item \textit{link to the same repository}: the referenced paper has a link back to the GitHub repository that contains a reference to that paper.
\item \textit{404}: the referenced paper has a link to a software repository, but the repository cannot be accessed.
\item \textit{no link}: the referenced paper does not have a link to a software repository.
\end{itemize}

 \begin{figure}[]
    \centering

\begin{subfigure}{\linewidth}
\centering
\caption{The number of academic papers in the top SE venues.}
\label{tab:rq2-sample}
\scalebox{0.70}{
\begin{tabular}{@{}lrr@{}}
\toprule
             & \textbf{\# papers} & \textbf{sample size} \\ \midrule
International Conference on Software Engineering         & 1,398              & 302                  \\
ACM SIGSOFT Symposium on the Foundations of Software Engineering          & 1,144              & 288                  \\
IEEE/ACM International Conference on Automated Software Engineering          & 1,233              & 293                  \\
Journal of Systems and Software          & 2,497              & 333                  \\
Information and Software Technology          & 1,488              & 305                  \\
Empirical Software Engineering         & 733                & 252                  \\
IEEE Transactions on Software Engineering          & 792                & 259                  \\ \midrule
\textbf{sum} & \textbf{9,285}     & \textbf{2,032}       \\ \bottomrule
\end{tabular}
}
\end{subfigure}
\hfill
\begin{subfigure}{\linewidth}
\centering
\includegraphics[width=\linewidth]{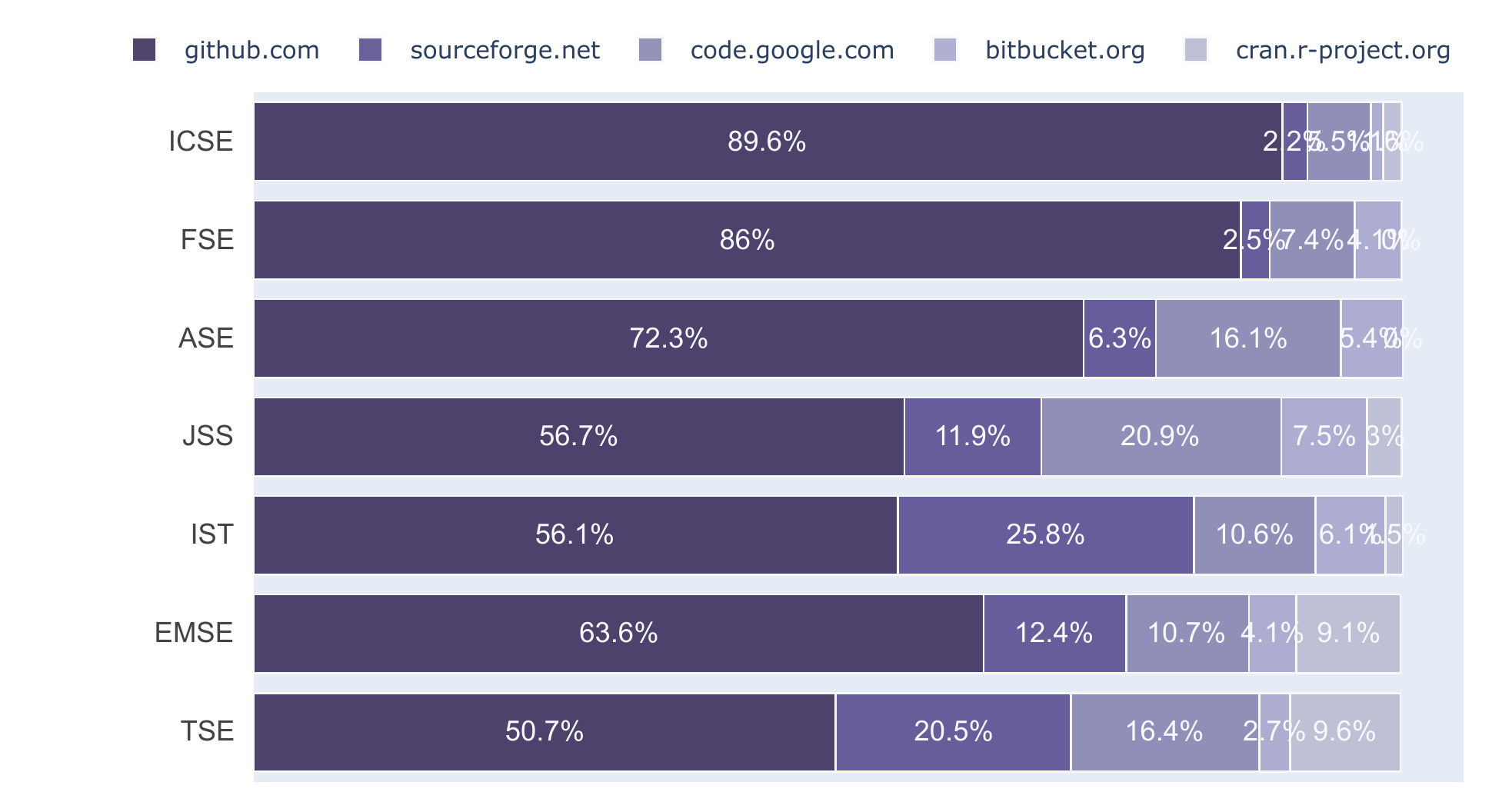}
\caption{Proportion of the top-5 most referenced software repository domains shared by the top SE venues.}
\label{fig:domain-freq}
\end{subfigure}
\hfill
\begin{subfigure}{\linewidth}
\centering
\caption{Frequency of link target types.}
\label{tab:rq24-links}
\scalebox{0.70}{
\begin{tabular}{@{}lrr@{}}
\toprule
                      & \multicolumn{2}{c}{\textbf{\# links}} \\ \midrule
software repositories & 990               & 60.3\%            \\
issues                & 193               & 11.8\%            \\
data repositories     & 100               & 6.1\%             \\
pull requests         & 63                & 3.8\%             \\
documentation         & 60                & 3.7\%             \\
GitHub home page      & 55                & 3.4\%             \\
commits               & 50                & 3.0\%             \\
user profile page     & 33                & 2.0\%             \\
source code           & 25                & 1.5\%             \\
files                 & 22                & 1.3\%             \\
GitHub blogs          & 11                & 0.7\%             \\
GitHub search pages   & 4                 & 0.2\%             \\
GitHub Apps           & 2                 & 0.1\%             \\
closed repositories   & 5                 & 0.3\%             \\
404 not found         & 28                & 1.7\%             \\ \midrule
\textbf{sum}          & \textbf{1,641}    & \textbf{100\%}    \\ \bottomrule
\end{tabular}
}

\end{subfigure}
\caption{Analysis of links from papers for RQ2.4.}
\end{figure}

Figure~\ref{fig:rq2.4} shows the results of our qualitative analysis for the 339 public access academic papers.
The results show that more than half of the academic papers referenced in the GitHub README.mds do not trace back to the GitHub repositories.
29\% of the referenced academic papers publish a link to their official repositories, and only 0.9\% of the referenced academic papers provide links that cannot be accessed (404 Not Found).
Considering ``official'' relationships only, 62 of the 136 academic papers cited by official repositories reference back to them, while 57 (42\%) papers do not trace back to the repositories.

We complement our results by also tracing papers that have not been collected through GitHub and reference a repository.
As such, we collect a sample of papers from the top 7 software engineering (SE) venues\footnote{ICSE, FSE, ASE, TSE, EMSE, JSS, and IST which is based on Google Scholar's ranking \cite{Software42:online}.} (with a confidence level of 95\% and a confidence interval of 5\% per venue), published between when GitHub was founded (2008) and 2020. 
From the sample of 2,032 papers, we find that ``github.com'' (1,641 link URLs) is the most referenced software repository domains, followed by ``sourceforge.net'' (106 link URLs), ``code.google.com'' (101 link URLs), ``cran-r-project.org'' (42 link URLs), and ``bitbucket.org'' (41 link URLs).
The number of papers and corresponding sample sizes for each venue is shown in Table~\ref{tab:rq2-sample}.

Figure \ref{fig:domain-freq} shows that for the top-5 most referenced software repository domains, more than half (55\%) of the ICSE papers in our representative sample contain links to those software repository domains, followed by FSE (39\%) and EMSE (38\%).
Furthermore, ``github.com'' is the most referenced domain across the top venues, especially in ICSE, where GitHub accounts for 89\% of link URLs reference to the 5 most referenced software repository domains.

Table \ref{tab:rq24-links} shows the diversity of target types of links to ``github.com''. 
Using regular expressions on the link URL and cross-checking the GitHub API, we were able to use a semi-automatic method to identify the type of GitHub link.
From the collected 1641 links, we find that only 60\% (990) of the links actually reference software repositories (top-5 most popular programming languages: Java (389), Python(174), JavaScript (73), C++ (63), and C (63)). Aside from the links to software repositories, 193 links reference issues, 100 links reference data repositories, 66 links reference pull requests, among others.

\begin{table}
\caption{Top 10 most referenced papers (i.e., In-degree) derived from our network of 20,278 GitHub repositories (as of September 2020). Interestingly, these papers turn out to be highly-cited.}
\centering
\scalebox{0.65}{
\begin{tabular}{llp{23mm}lp{68mm}r}
\toprule
\textbf{rank} & \textbf{In-degree} & \textbf{\# repo type} & \textbf{identifier} & \textbf{paper}  & \textbf{\# referenced by} \\ 
 & \multicolumn{2}{c}{\textbf{}} &  &  & \textbf{academia} \\
\midrule
1 & 207 & Py:201, JS:4, C++:2 & 1512.03385 & \citet{He_2016}: \textbf{Deep Residual Learning for Image Recognition} (2016) & 55,095 \\
2 & 189 & Py:181, JS:5, Java:2, C++:1 & 1409.1556 & \citet{simonyan2014deep}: \textbf{Very Deep Convolutional Networks for Large-Scale Image Recognition} (2014) & 43,848 \\
3 & 170 & Py:160, C++:9, C:1 & 1511.06434 & \citet{radford2015unsupervised}: \textbf{Unsupervised Representation Learning with Deep Convolutional Generative Adversarial Networks} (2015) & 7,112 \\
4 & 159 & Py:152, C++:4, JS:2, Java:1 & 1508.06576 & \citet{DBLP:journals/corr/GatysEB15a}: \textbf{A Neural Algorithm of Artistic Style} (2015) & 1,395 \\
5 & 139 & Py:135, C++:3, Java:1 & 1602.01783 & \citet{mnih2016asynchronous}: \textbf{Asynchronous Methods for Deep Reinforcement Learning} (2016) & 3,683 \\
6 & 136 & Py:135, JS:1 & 1408.5882 & \citet{DBLP:Kim14f}: \textbf{Convolutional Neural Networks for Sentence Classification} (2014)
& 8,386 \\
7 & 135 & Py:133, C:1, JS:1 & 1706.03762 & \citet{DBLP:VaswaniSPUJGKP17}: \textbf{Attention Is All You Need} (2017) & 11,727 \\
8 & 114 & Py:113, C++:1 & 1509.06461 & \citet{hasselt2015deep}: \textbf{Deep Reinforcement Learning with Double Q-learning
} (2015) & 2,405 \\
9 & 110 & Py:107, C++:2, JS:1 & 1608.06993 & \citet{huang2016densely}: \textbf{Densely Connected Convolutional Networks} (2016) & 10,967 \\
9 & 110 & Py:104, C++:3, C:1, Java:1, JS:1 & 1703.0687 & \citet{kokot2017faster}: \textbf{Even faster sorting of (not only) integers} (2017) & 7 \\
                           \bottomrule
\end{tabular}%
}
\label{tab:arxiv}
\end{table}

To answer RQ2.4, our findings indicate that \textit{a large number of academic papers do not reference back to the software repositories even though being referenced by the official repositories.
By analyzing papers from the top-tier SE venues, we find that many academic papers do not reference (i.e., hyperlinks) back to software repositories, but when they do, they usually link back to a GitHub domain.
This leads to an undesirable gap between the academic papers and the OSS community.} 

\paragraph{(RQ2.5) Most Referenced arXiv Papers}
To answer RQ2.5 and RQ2.6, we identify which papers are referenced the most from GitHub repositories.
We conducted a case study with the subset of links pointing to arXiv.org, one of the largest public access platforms with around 1.6 million e-prints of academic papers.
To identify arXiv papers, similar to the pattern matching used in the data collection phase (see Section \ref{subsec:data_collection}), we identified all arxiv.org links that include a unique identifier (i.e., https://arxiv.org/*/xxx.xxx).
We then use network analysis to answer RQ2.5. 
Particularly, we generated a bipartite network using the repositories and the papers as two different types of nodes, and the links between them as directed edges (from repositories to papers). 
The network constructed from the 20,278 repositories contains 20,373 nodes (i.e., repository: 58.8\% and paper: 41.2\%). 
Using this network, we located the most referenced papers by using the in-degree measure of the paper nodes (representing the number of repositories in which one paper is referenced).

Table \ref{tab:arxiv} shows the 10 most referenced academic paper from GitHub README.mds.
All the top academic papers are from the fields of deep learning and computer vision, and were published between 2014-2016, which is arguably SOTA.
The table also shows these most referenced papers are highly-cited papers in academia and later published in distinguished AI conferences. 
Although the paper \#10 has only 7 citations, the paper has more than 800 downloads as demonstrated in the SpringerLink \footnote{\url{https://link.springer.com/chapter/10.1007/978-3-319-67792-7\_47}} website.
This is a rare case for papers that have more practical impact than in academia.
To answer RQ2.5, our findings indicate that \textit{the most referenced papers in GitHub repositories are highly-cited in academia as well.} 

\paragraph{(RQ2.6) Programming Language Diversity of Repositories that Reference arXiv Papers}
Our approach for answering RQ2.6 involves analysis of the connectivity of the nodes from the same bipartite network that was constructed for answering RQ2.5.
We analyze the number of incoming edges from a repository to a paper (i.e., in-degree).

Table \ref{tab:arxiv} and Table \ref{tab:RQ26c} describe the diversity of programming languages in our network.
The results are broken into two analyses: (i) the diversity of repositories that only reference a single paper and (ii) the diversity of repositories that reference the top-10 most influential papers in our network.
Our finding first show that in both cases, most of the repositories are written in the Python language.
This is not surprising, as Table \ref{tab:collected_repo} clearly shows a significant number of repositories are Python repositories (i.e., 14,073 out of 20,278).
Our network has a long tail, with almost 50\% of the papers being referenced by a single repository.
It is interesting to note that C++ is clearly the second most popular programming language, with 747 repositories. 
Taking a look at the top-10 most referenced papers, Python also dominates the repositories which link to these papers, with C, C++, Java, and JavaScript repositories containing less references.
To answer RQ2.6, our findings indicate that \textit{although almost 50\% of papers are only referenced by a single repository, the most referenced papers can be referenced in software repositories that implement software written in different programming languages.}

\begin{table}[]
\centering
\caption{Count of papers that are referenced by a single repository (in-degree = 1).}
\centering
\scalebox{0.80}{
\begin{tabular}{@{}lrr@{}}
\toprule
\textbf{programming language} & \multicolumn{2}{r}{\textbf{\#repositories}} \\ \midrule
Python               & 4,111            & 72\%            \\
C++                  & 747              & 13\%            \\
JavaScript           & 351              & 6\%             \\
C                    & 266              & 4.6\%           \\
Java                 & 232              & 4\%             \\
Ruby                 & 17               & 0.3\%           \\
PHP                  & 6                & 0.1\%           \\ \midrule
\textbf{sum}                  & \textbf{5,730}            & \textbf{100\%}           \\ \bottomrule
\end{tabular}
}
\label{tab:RQ26c}
\end{table}

\begin{tcolorbox}
\textbf{Summary}: 
In terms of traceability, there is a gap between software repositories and papers. We find that more than half of these referenced papers do not link back to any repository. By analyzing papers from the top-tier SE venues, we find that a large number of academic papers do not reference (i.e., hyperlinks) software repositories, but when they do, they usually link back to a GitHub software repository. Taking a look at arXiv papers, we find that the most referenced papers are also highly-cited in academia, and are referenced by software repositories written in different programming languages.
\end{tcolorbox}

\subsection{Evolution (RQ3)}
To understand how the links evolve (RQ3), we investigated the revision histories of repositories in the sample of 339 repositories that have links to public access academic papers. Our evolutionary analysis is summarized from the two perspective of the papers (RQ3.1) and README.mds (RQ3.2). We again conducted a qualitative analysis of our statistically representative sample, this time focusing on the evolution of the papers and README.mds. We utilized the same approach as in RQ2 to design and validate the coding guides by asking four authors to independently code 30 cases.

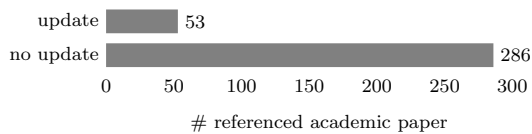
\begin{figure}[t]
        \centering
        \begin{footnotesize}
        \begin{tikzpicture}[scale=0.9]
        \begin{axis}[
            align=center,
            y=0.5cm,
            x=0.02cm,
            enlarge y limits={abs=0.25cm},
            symbolic y coords={no update,update},
            axis line style={opacity=0},
            major tick style={draw=none},
            ytick=data,
            xmin = 0,
            xlabel = \# referenced academic paper,
            nodes near coords,
            nodes near coords align={horizontal},
            point meta=rawx
        ]
        \addplot[xbar,fill=gray,draw=none] coordinates {
            (53,update)
            (286,no update)
        };
        \end{axis}
        \end{tikzpicture}
        \end{footnotesize}
        \caption{Answering RQ3.1, our findings indicate that the majority of referenced academic papers (84\%) do not get any update after being referenced.}
        \label{fig:rq3.1}
    \end{figure}

\paragraph{{(RQ3.1) Paper evolution}}
We used the timestamp of the papers and commits of the README.mds to investigate whether the papers were updated after being referenced, e.g., we consider a paper has an update if a new version of the paper was published on arxiv.org or the paper was extended to a journal at a later point in time compared to when the paper is being referenced. Since paper content cannot be changed after publication, we did not consider the changes in the contents of the referenced papers for this analysis. The following list shows two codes that we used to annotate paper evolution.
The kappa agreement was 0.87 or ``almost perfect''~\cite{viera2005understanding}.

\begin{itemize}
\item \textit{no update}: the referenced academic paper did not get updated after being referenced.
\item \textit{update}: the referenced academic paper was updated after being referenced.
\end{itemize}

Figure \ref{fig:rq3.1} summarizes the results of this analysis.
We found that 286 (84\%) of the referenced academic papers in our representative sample had not been updated at all. A significant minority of 16\% of the referenced papers were revised.
To answer RQ3.1, our findings indicate that \textit{most academic papers that are being referenced in GitHub README.mds did not undergo any changes after they were referenced. However, a significant minority of papers were updated after the corresponding reference in the GitHub README.md was added.}

\paragraph{{(RQ3.2) README.md evolution}}
To answer this sub-question, we also focused on the timestamp of README.md commits to investigate changes in the links to academic papers in GitHub README.mds. We did not focus on actual source code changes for this analysis. The following list shows codes that emerged from our analysis of the evolution of the links to academic papers in GitHub README.mds, along with a short description.
The kappa agreement was 0.91 or ``almost perfect''~\cite{viera2005understanding}.

\begin{itemize}
\item \textit{different paper}: the README.md referenced a different academic paper before. 
\item \textit{changes to metadata}: the meta data of the paper was changed, e.g., from ``to appear'' to the publication information.
\item \textit{link changes}: a link to the referenced academic paper was updated, e.g., from a pdf stored on GitHub to an arXiv or DOI link.
\item \textit{no update}: the reference had not evolved.
\end{itemize}

 \begin{figure}[t]
        \centering
        \begin{footnotesize}
        \begin{tikzpicture}[scale=0.9]
        \begin{axis}[
            align=center,
            y=0.5cm,
            x=0.02cm,
            enlarge y limits={abs=0.25cm},
            symbolic y coords={no update, different paper, changes to metadata, link changes},
            axis line style={opacity=0},
            major tick style={draw=none},
            ytick=data,
            xlabel = \# GitHub README.mds,
            xmin = 0,
            nodes near coords,
            nodes near coords align={horizontal},
            point meta=rawx
        ]
        \addplot[xbar,fill=gray,draw=none] coordinates {
            (6,different paper)
            (8,changes to metadata)
            (14,link changes)
            (311,no update)
        };
        \end{axis}
        \end{tikzpicture}
        \end{footnotesize}
        \caption{Answering RQ3.2, we show that up to 92\% of references to academic papers do not evolve in the README.mds.}
        \label{fig:rq3.2}
    \end{figure}
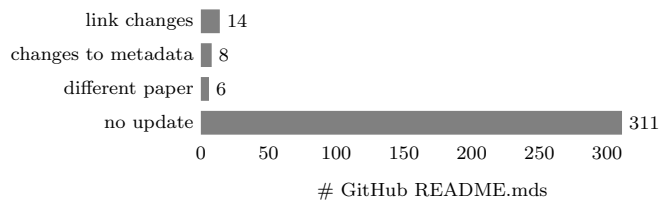

Figure \ref{fig:rq3.2} shows the number of changes to the references after they were first added to a GitHub README.md.
The result shows that 92\% of the references have not been updated.
Only 8\% of the links in the 339 README.mds had at least one change, with ``link changes'' being the most frequent (e.g., the link to a pre-print version of the paper superseded by a newer link to an official website).
The majority (11 of 14) of README.mds for which links to academic papers were updated are in GitHub official repositories, i.e., authors' ones.
While not all changes to a paper require updates to README.md linking to it, especially if the link does not change, we expect that some changes to a paper will trigger corresponding changes in a repository.
To answer RQ3.2, our findings indicate that \textit{a large number of references to academic papers in GitHub README.mds do not attract any changes, but a small number of links are updated.}

\begin{tcolorbox}
\textbf{Summary}: 
Evolution is rare. We find that while some academic papers that were referenced in GitHub README.mds change, most of the papers do not. Furthermore, these changes are likely not reflected in repositories.
\end{tcolorbox}

\section{Limitations and Implications}
\label{sec:recommendations}

Based on our findings, we identified the following limitations and highlight the implications of the study.

\subsection{Limitations}
The first challenge is related to improving our search matching pattern.
The downside of our search string is the bias towards arXiv and DOI links, which does not extend to other academic public access (i.e., PeerJ\footnote{https://peerj.com/}) and closed access paper portals (i.e., IEEE Xplore\footnote{https://ieeexplore.ieee.org/}, ACM digital libraries\footnote{https://dl.acm.org/}).
Our results show that the ML community has a strong presence in GitHub.
Immediate future work would be to extend the pattern for generalization of our findings.

The second challenge raised by this study is to understand the indirect impact of an academia paper on the repository. Although not provided in this study, potential future work could be exploration of what kinds of software changes are made after a paper is super-seeded as SOTA and vice-versa. Furthermore, in RQ2.5, we show the direct impact of a paper in academia (i.e., citations), however, we did not explore the impact of the GitHub repository in terms of forks and stars. This could also quantify how much impact the software implementation has in practice (e.g., an open-source library implements and references an algorithm proposed in a paper while other libraries build on this library without referencing the initial paper). In this work, we only select one paper per repository, however, we find that a repository may reference many papers. Based on the code changes, a potential future avenue is to explore the relationship between the papers and the different locations in the source code. Finally, feedback from developers could also bring insights into how to bridge the gap between academic papers and the software implementations.

\subsection{Implications}
We now discuss the implications for software engineering researchers, academic authors, and GitHub users in both academic and industrial fields. Based on our findings, we are able to provide practical and actionable implications:

\paragraph{For SE researchers}
Our results indicate that while referencing from GitHub repositories to academic papers is not that common, the references that do exist pre-dominantly go to public access papers. One of the main reasons is that arXiv is able to create a permanent link which guarantees that a publication is considered a permanent part of the scientific record and may not be removed by user intervention \cite{arXivorg85:online}. There are also other services like Zenodo and PeerJ as locations for storing a paper. We encourage software engineering researchers to continue to make their work public access for practitioners, ideally in public access portals like arXiv, which is better than papers stored on private websites since a private website address does not guarantee that a publication is available at a particular URL, especially there is a possibility that authors remove their work from the website \cite{swcitaion}.
Top software engineering venues, i.e., the 43rd International Conference on Software Engineering (ICSE 2021) and Empirical Software Engineering journal (EMSE), also encourage contributing researchers to disclose their research materials and self-archive the pre- and post-prints, especially in open and preserved repositories, for making their work more transparent \cite{10.1007/s10664-019-09712-x, Heumller2020PublishOP}.

\paragraph{For academic authors}
We find that many academic papers do not link back to any repository while a few academic papers publish a link to authors' repositories in GitHub. We encourage academic authors to include bi-directional links wherever possible, but also envision automated tool support to help achieve this. This is especially important for accountability of the scientific results \cite{norawdata20}, and increasing impact and innovation as researchers see the fruits of their scientific findings. We believe that tool support for augmenting academic papers with the corresponding references to GitHub repositories is needed. Implementing this could be fairly straight-forward, either through a browser plugin or through a bot which leaves comments on the arXiv homepage of a paper (or another platform) whenever a relevant repository is found.

\paragraph{For GitHub developers}
We show that GitHub is being used as a platform for sharing research outputs, eventhough there are other more suitable platforms (i.e., kaggle\footnote{https://www.kaggle.com/}, zenodo\footnote{https://zenodo.org/})).
Furthermore, our results indicate that while both the repositories and the papers are rarely updated, some papers get superseded as the SOTA after being referenced.
We believe that tool support could notify developers when the SOTA in related research gets updated.
This could be a pull request that notifies when there is the cutting edge research that supersedes the existing work or when switching over to an official version from a preprint version.

\section{Threats to Validity}
\label{sec:threats_to_validity}
Regarding threats to \textit{internal validity}, it is possible to introduce bias through our qualitative analysis. We mitigate these threats by reporting the agreement amongst four authors using the Cohen's kappa.
An example was the resolution of all disagreements from the initial coding guide in RQ2.1.
In this case, all four authors discussed to refine the coding guide and re-coded until the question achieved high kappa agreement (more than 0.8).

Threats to the \textit{construct validity} exist in our approach to link identification, since we identified links in README.mds based on pattern matching. 
Our search string matching patterns does not extend to (1) PDF extension, (2) All \BibTeX{}, and (3) other patterns: (1) \texttt{`+\textbackslash.pdf\$'}, (2) texttt{`@book\{'}, \texttt{`@booklet\{'}, \texttt{`@conference\{'}, \texttt{`@inbook\{'}, \texttt{`@incollection\{'}, \texttt{`@manual\{'}, \texttt{`@mastersthesis\{'}, \texttt{`@phdthesis\{'}, \texttt{`@proceedings\{'}, \texttt{`@techreport\{'}, \texttt{`@unpublished\{'}, (3) \texttt{`peerj.com'}, \texttt{`dl.acm.org'}, \texttt{`ieeexplore.ieee.org'}, \texttt{`springer.com'}, \texttt{`elsevier.com'}, \texttt{`computer.org'}, \texttt{`researchgate.net'}, \texttt{`semanticscholar.org'}, with the key reasons of introducing more noise (i.e., links to other documents or html pages) as shown in Table \ref{tab:other-patterns}, while the patterns may be biased towards machine learning repositories caused by including arXiv in our matching pattern.
We are confident that our search string matching patterns achieve low false positives since they exactly match references  in  explicit \BibTeX{} formats and explicit links pointing to DOI for academic papers and arXiv.org. With the findings on the additional analysis on expanded patterns, we ensure that our matching patterns are still able to cover other patterns.

Threats to the \textit{external validity} exist in our repository preparation and sampling. Although we analyzed a large amount of repositories on GitHub, we cannot generalize our findings to industry nor open source repositories in general since some open source repositories are hosted outside of GitHub, e.g., on GitLab or private servers. To mitigate the threats related to sampling selection, we focused on those containing code written in at least one of seven popular programming languages. We selected 500 repositories from each language for our preliminary analysis, and drew a statistically representative sample across the seven languages for the qualitative parts of our work. Focusing on different languages or not using programming languages as filter at all might have led to different results. We implemented these filters to ensure that the repositories considered in this work contained source code and that the repositories were not used exclusively for academic paper writing, for example.

\section{Related work}
\label{sec:related_work}
Our work is situated in the literature on (1) traceability among different scientific artifacts and (2) the references in open source software documentation.

\paragraph{Traceability among Scientific Artifacts}
How to increase the incentives and integration for scientific software development has attracted the attention of researchers who are studying software engineering, informatics, and many fields that rely on scientific computing~\cite{hannay2009scientists, howison2013incentives}. Open source has been considered as one of the most common solutions~\cite{prlic2012ten}. The rapid progress in machine learning research recently has accelerated the progress of open scientific software development. To promote bench-marking or adoption, there is a pressing needs to encourage the practice of releasing multiple scientific research artifacts (e.g. papers, source code, and datasets) and making explicit trace links among them~\cite{Drummond2009, gibney2020ai}. Community-driven initiatives such as \texttt{Papers With Code} \cite{PapersWi17:online} and \texttt{RedditSOTA} \cite{RedditSo59:online} are examples of direct responses to addressing such need. While those third-party platforms can serve as centralized places to find SOTA results on popular machine learning problems, they require additional submission and maintenance not necessarily from the original authors of the work. 

As far as the authors are aware, there is no existing work that systematically examines the characteristics of trace links to academic papers directly embedded in the open-source repositories, especially how they are created and maintained. The most relevant work from the software engineering research community is the one conducted by Milewicz et. al.~\cite{MilewiczMSR19} in which the role of the contributors in open-source scientific software repositories was investigated. Their empirical analysis on seven open-source scientific software repositories and the complementary survey study revealed that the tenure of the contributors is the decisive factor to determine their role in the repository. While part of their target repositories were selected from GitHub, they used the tags of the repository to identify scientific software repositories. They did not inspect any links to the other scientific artifacts such as papers to validate their repositories. 

Recently, the impact of open scientific software has been gone beyond the scope of scientific computing in academia. Braiek et al.~investigated the ecosystem of popular open-source machine learning frameworks~\cite{BraiekMSR18}. They found that many such frameworks were initially created by the academic communities but were then taken over by the support from companies. Their analysis of the composition of the development team for those repositories revealed the tendency that professional researchers normally contribute equally with engineers for company-driven ML repositories, while professional and academic researchers contribute mostly for the community-driven repositories. Their work, however, did not consider if any academic papers were accompanied by the development of those repositories, nor if any trace links exist in those repositories.

\paragraph{References in OSS Documentation}
GitHub README.mds were studied systematically by Prana et al.~as one specific type of software documentation in their recent work~\cite{prana2019categorizing}. Their qualitative study revealed that the content in the README.mds can be categorized into eight groups among which the second most frequent categorize was particularly about links to other resources. This result indicates that project contributors consider links as an important instrument to provide further details about their projects.
\citet{DBLP:IkedaIKM19} performed a similar study to evaluate README.md contents.
Inspired by these works, we initiated our investigation on the relationship between academic papers and the scientific projects hosted on GitHub through the reference links embedded in their project README.mds.

Hata et al.~investigated the characteristics of the reference links written in source code comments for open source projects, in particular, the purpose of those links and how they are evolved and decayed~\cite{10.1109/ICSE.2019.00123}. 19 different kinds of links have emerged from their analysis, one of which was the links to academic papers. They also divided the target of the links based on the domain name of the link URL and found that the links to academic papers normally belong to the sometimes or rarely linked domains. They have also found dead links for research papers indicating the problem with maintaining those links in source code.
Otherwise, we found some dead links in papers (see Section 3.2).
In this work, we zoom in on the maintenance and evolution practice of the links to academic papers in the README.mds using the version control system. Furthermore, we also looked at the evolution of scientific work after the link was created. These analyses gave us a comprehensive understanding of the practices and challenges for synchronizing scientific artifacts.

\section{Conclusion}
\label{sec:conclusion}
To understand public access, traceability, and evolution of links to academic papers in GitHub repositories, we conducted (i) a quantitative study of around 20 thousand links from GitHub README.mds in 10,343,311 GitHub repositories to establish the frequency of these links; (ii) a qualitative and quantitative study of a sample of 344 README.mds to determine public access; (iii) a qualitative study to trace and analyze the relationships between the academic papers and the repositories that reference them; and (iv) a quantitative study to determine the evolution of academic papers and updates to the repository.

Our work has shown that referenced academic papers are indeed public access, with most referenced papers being regarded just as influential for the academic community. Based on this work in documentation files, insights that are revealed from the study open many avenues for future work: investigate strategies to promote public access and reproducibility, further study to understand how research accelerates through the alignment of SOTA between research and practice, tool support for recommendation of related academic papers and repositories, study the nature of repositories that link to academic papers, investigate the indirect impact of academic papers (e.g., an open-source library implements and references an algorithm proposed in a paper while other libraries build on this library without referencing the initial paper), survey repository owners whether they are aware of changes in paper updates, and further understand impacts of the paper updates on repositories, to name a few.

\section{Acknowledgement}
This work is supported by JSPS KAKENHI Grant Number 16H05857, 18H04094, 20K19774, and 20H05706 as well as by the Australian Research Council’s Discovery Early Career Researcher Award (DECRA) funding scheme (DE180100153).

\bibliographystyle{elsarticle-num-names}
\bibliography{bibliography}

\end{sloppy}
\end{document}
\endinput